# Title: Strong, Mode-Selective Evxciton-Photon Coupling Driven by Polariton Scattering in the $Mo_2$ Complexes at Ambient Conditions

Miao Meng, Ying Ning Tan, Zi Cong He, Yuli Zhou, Chun Y. Liu*

**Affiliations:**

*Department of Chemistry, College of Chemistry and Materials Science, Jinan University, 601 Huang–Pu Avenue West, Guangzhou 510632, China*

Correspondence to: tcyliu@jnu.edu.cn

**Abstract**

Quadruply bonded $Mo_2$ complexes provide a distinctive molecular platform in which multiple two-level electronic transitions interact with quantized scattering modes under ambient conditions. Here we show that, in the $Mo_2$ complexes, the intrinsic photonic modes selectively couple to molecular excitations, including the characteristic δ→δ*, ligand to metal charge transfer (LMCT) and metal to ligand charge transfer (MLCT) transitions, to form the well-resolved exciton–photon hybrid states. By combining steady-state absorption, resonance fluorescence, and ultrafast transient spectroscopies, we identify distinct polaritonic branches associated with these electronic manifolds, with coupling strengths spanning the strong and ultrastrong regimes ($g/\omega_0$ up to 0.1). Mode-selective coupling accounts for the pronounced spectral reorganization for the singly oxidized complexes, including the emergence of absorption valleys and sidebands of the associated resonances in the steady-state spectra, characteristic polaritonic emissions in the photoluminescence spectra, and long-lived low-energy hybrid states in the transient spectra. These results support the picture that the $Mo_2$ unit functions as an integrated molecular resonator whose intrinsic quantized field selectively drives and redistributes molecular excitations. This work strengthens the emerging view of bonded dimetal complexes as ambient-condition molecular quantum systems and provides a chemically defined platform for exploring polaritonic chemistry.

## Introduction

Strong light-matter interaction is traditionally realized by coupling a two-level transition of an atom or molecule to a discrete photonic mode of an optical cavity. In such a system, the cavity field is quantized, and the corresponding mode energies scale with the photon number state $|n\rangle$ according to the Jaynes-Cummings framework, giving rise to the characteristic $\sqrt{n+1}$ dependence of the dressed-state splitting.[1,2] [3,4,5] This quantized light-matter interaction was first established experimentally for Rydberg atoms coupled to a high-$Q$ cavity under cryogenic conditions.[6] Under resonant coupling, the electronic excitation and the cavity mode exchange energy coherently through Rabi oscillation between the dressed states. These behaviors define a central paradigm in quantum optics and have motivated broad interest in quantum communication and information processing,[7,8] and control of chemical reaction and electron (energy) transfer processes.[9,10,11]

In our recent studies of dimetal complexes,[12,13,14,15] we found that radiative decay of excited $M_2$ systems ($M_2 = Ni_2$ and $Mo_2$) in dilute solution produces discrete emissions in the photoluminescence spectra, with transition energies that follow square-root scaling with an integer number $N$. This ladder-like spectral structure is consistent with resonance fluorescence in the strong-coupling regime for both single ($N = 1$) and few-molecule ($N = 2, 3$) systems, where each molecule behaves as an effective Jaynes-Cummings unit, with the M $\rightarrow$ M charge-transfer excitation resonantly coupled to the photonic mode of the scattering field of the $M_2$ core.[13,14,16,17] Spectral analysis further suggested that the scattering field is quantized through collective excitation of $N$ molecules.[13] The polaritonic mode frequencies correspond to the dressed states, scaled as $\omega_0 \pm (N\sqrt{N}\Omega_1)/2$ ($\Omega_1$ = 1380 cm$^{-1}$) for the Rabi splitting states and $\omega_0 \pm N\Omega_1'$ ($\Omega_1'$ = 1450 cm$^{-1}$) for the Mollow-triplet sideband states (Figure 1A). The collective coupling behavior of the $N$-molecule system is in accordance to the Jaynes-Cummings physics,[1,18] but distinct from that predicted from the Tavis-Cummings model.[19] Within this framework, the molecule-field interaction in the $M_2$ system converts incident classical light into nonclassical scattering modes, enabling polaritonic responses in free space under ambient conditions. [14] An important implication of this cavity-free platform is that exciton-photon coupling can become mode selective, such that a given molecular electronic manifold couples preferentially to a discrete subset of intrinsic field modes. Demonstrating such mode-selective coupling would therefore provide strong support for the integrated emitter-resonator picture and for field quantization in this molecular quantum system. [13,14,15]

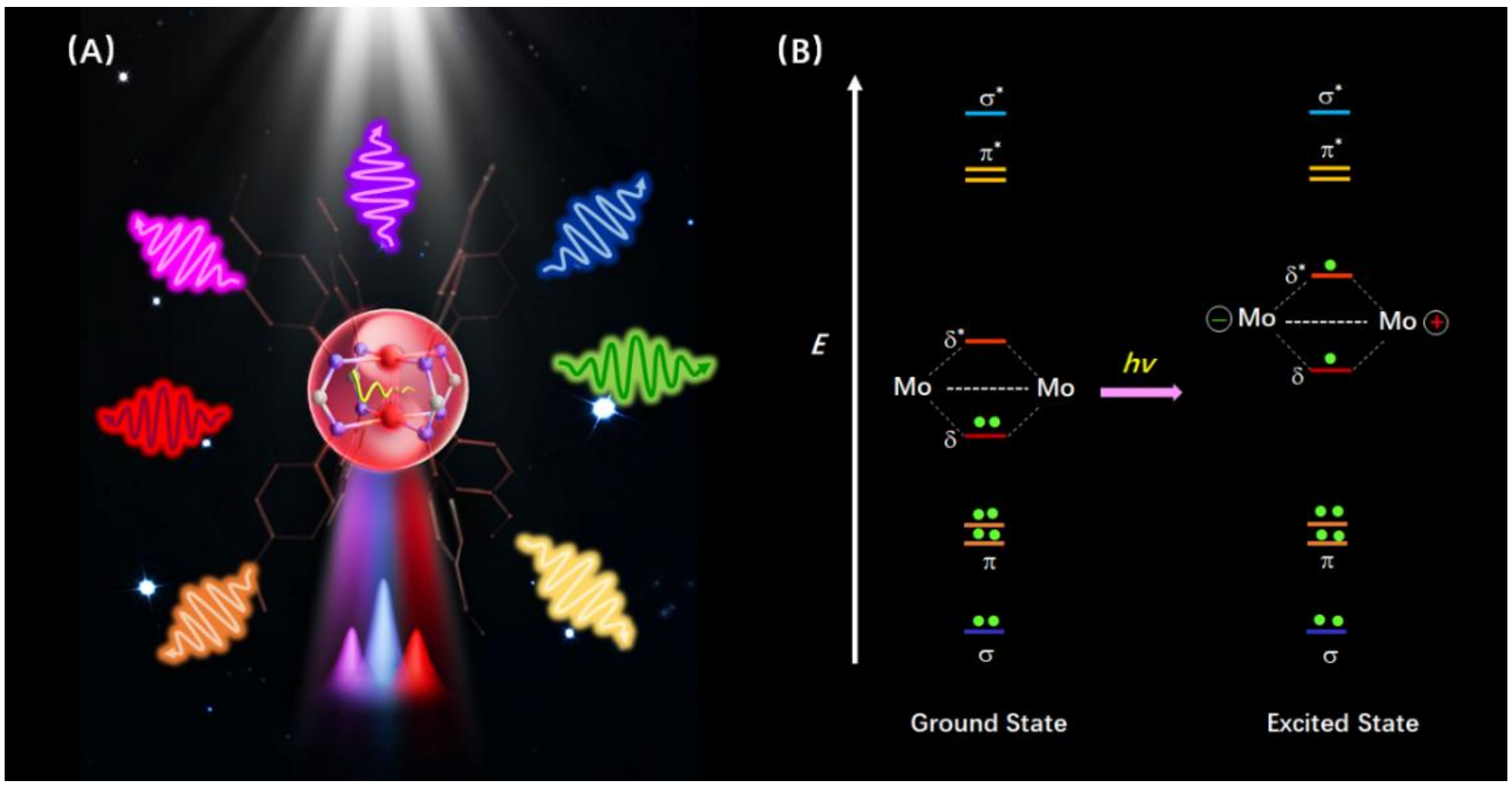

**Figure 1.** (A) Quantization of the scattering field of the $Mo_2(DAniF)_4$ complex under visible-light excitation, generating discrete nonclassical photonic modes over a broad wavelength range. (B) Schematic Illustration of the ground-state electronic configuration of the $Mo_2$ unit, $\sigma^2\pi^4\delta^2$, and the corresponding excited zwitterionic states. The characteric two-level excitations of the $Mo_2$ complexes include the $\delta \rightarrow \delta^*$ and two ligand to metal charge transfer (LMCT) transitions, ligand $\rightarrow \delta^*$ and ligand $\rightarrow \pi^*$.

Quadruply bonded dimolybdenum complexes are particularly well suited for this objective because their electronic structure provides several well-defined two-level excitations that can be matched energetically to the intrinsic scattering modes. Formation of the Mo–Mo quadruple bond lifts the *d*-orbital degeneracy of the isolated metal centers and gives rise to the ground-state configuration $\sigma^2\pi^4\delta^2$, together with the characteristic $\delta \rightarrow \delta^*$transition that is spectroscopically accessible in the visible region (Figure 1B).[20] In addition to this metal-centered excitation, paddle-wheel $Mo_2$ complexes display higher-energy ligand to metal charge transfer (LMCT) transitions involving donor-atom orbitals of the supporting ligands and the $Mo_2$-based antibonding $\pi^*$or $\delta^*$orbitals (Figure 2).[21,22] In complexes bearing conjugated ancillary ligands, a low-lying metal to ligand charge transfer (MLCT) excitation can also arise through promotion of a $\delta$ electron into a ligand-centered $\pi^*$ orbital (Figure 2C).[12] These electronic manifolds therefore provide a chemically defined set of two-level excitations spanning the near-UV and visible regions. Remarkably, the quantized scattering field of the $Mo_2$ system contains discrete modes whose energies closely match these transitions,[14] making $Mo_2$ complexes a favorable platform for validating the mode frequencis of the scattering field and for testing whether the intrinsic molecular field modes can selectively drive exciton-photon coupling under ambient conditions.

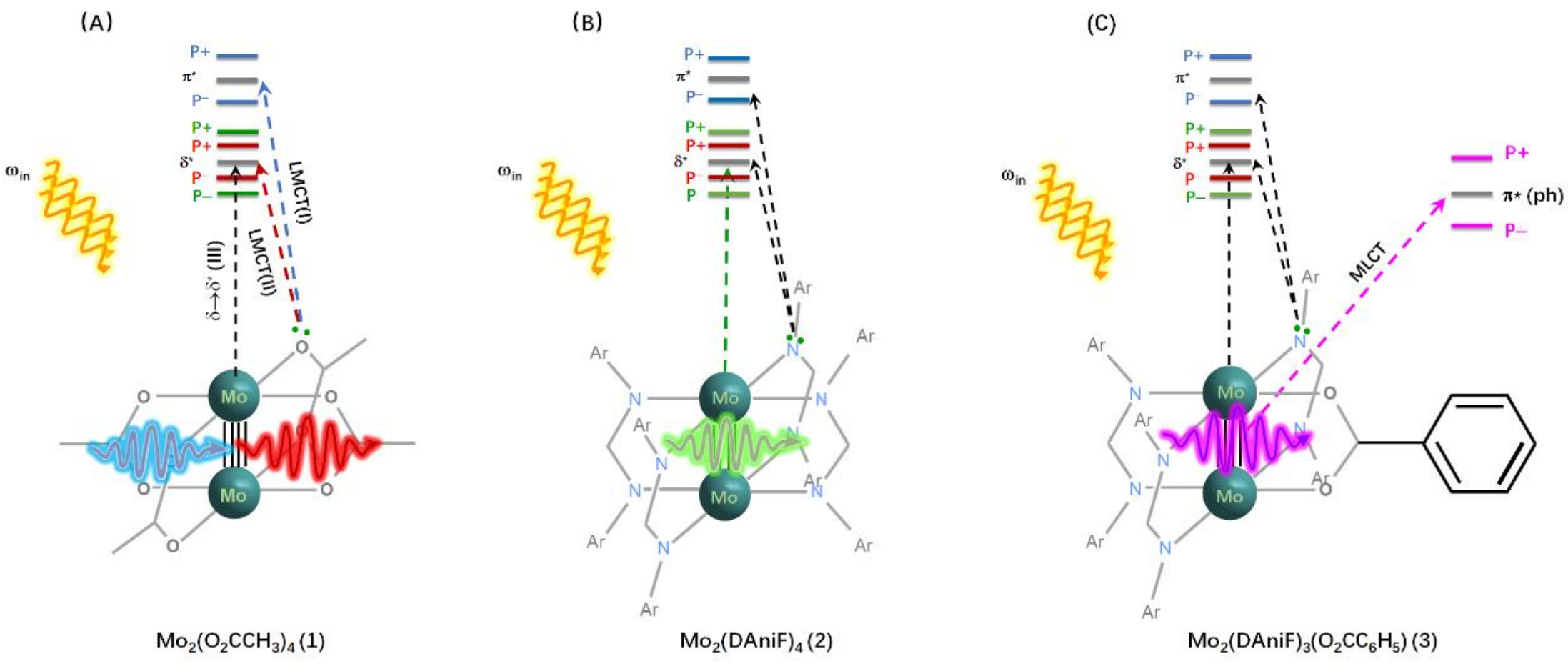

**Figure 2. Schematical illustration of the characteristic two-level transitions in the $Mo_2$ complexes and their Mode-selective cupling to intrinsic scattering-field modes, leading to vacuum Rabi splitting of the resonance.** (A) Resonant coupling of the LMCT transtions O $\rightarrow$ $\pi$* (**I**) and O $\rightarrow$ $\delta$* (**II**) to the photonic modes $\omega_{313}$ and $\omega_{392}$, respectively, for complex **1**. (B) Resonant coupling of the $\delta\rightarrow\delta$* transiton (**III**) to the scattering field mode $\omega_{443}$ for complex **2**. (C) The MLCT transition at 395 nm $\delta$ $\rightarrow$ $\pi$*(ph) is near-resonantly coupled to the scattering field mode $\omega_{392}$ in complex **3**. The transition and fiel mode are marked by the pink color icons. Colored icons denote the corresponding transitions and field modes.

In this work, we examine three $Mo_2$ complexes, $Mo_2(O_2CCH_3)_4$ (**1**), $Mo_2(DAniF)_4$ (**2**) (DAniF = *N*, *N'*-di(*p*-anisyl formamidinate) and $Mo_2(DAniF)_3(O_2CC_6H_5)$ (**3**), in the context of molecular excitation-scattering field coupling. These systems involve three characteristic transitions: LMCT (**I**), MLCT(**II**), δ→δ*(**III**), and MLCT for **3**. Our focus is on strong coupling between selected molecular excitations and individual intrinsic modes of the scattering field within a Jaynes-Cummings-type framework. The key experimental signature is the mode-governed redistribution of spectral weight into polaritonic branches that appear at well-defined wavelengths associated with the intrinsic field-mode set.[13,14] In this picture, coupling suppresses the original electronic absorption at resonance and generates split sidebands corresponding to transitions to the dressed states. Notably, in our previous studies,[12,14]these coupling signatures were resolved much more clearly and reproducibly in the singly oxidized $Mo_2$ complexes than in the corresponding neutral precursors, whose spectra are often broadened by overlapping electronic absorptions. This charge-state dependence of coupling strength cannot be understood simply as a conventional redox-induced shift of transition energies. Instead, it suggests that oxidation changes the nature of the relevant molecular excitons and thereby alters the pathway for scattering-driven exciton-photon coupling. In the present work, we analyze this contrast between the neutral and oxidized complexes in terms of the formation, binding, and coupling behavior of the associated excitons, and use this framework to explain the distinct optical responses of the cationic species.

## Results and Discussion

**Mode-selective coupling of the molecular transtions by the scattering field.** In dilute dichloromethane solution, the neutral $Mo_2$ complexes **1–3** display broad and partially overlapping absorptions arising from transitions **I–III** (Figures 2A and 2B), which complicates accurate identification of the individual transition energies (Figures 3A–3C). For the complexes examined here, the two LMCT transitions **I** and **II** occur in the near-UV region, around 310 nm and 390 nm,[23] respectively, whereas the δ→δ* transition (**III**) is obsesrved in a narrow range of 435-455 nm in the electronic spectra.[12,22,23,24] By contrast, the corresponding one-electron-oxidized species (Figure S1) exhibit stronger and more highly reorganized absorption features, with maxima shifted away from the electronic resonances of the neutral complexes. This spectral contrast cannot be accounted for by electronic structure changes alone, because previous calculations for **2** and **2**$^+$ showed broadly similar frontier-orbital patterns and comparable underlying transition energies for the neutral and oxidized molecules.[22] To resolve the difference more clearly, we analyze the differential spectra (ΔA) obtained by subtracting the molar absorptivity of each neutral complex from that of its oxidized counterpart over the same spectral range (Figures 3A–3C).[12,25] In all three cases, the oxidized species display more intense and more strongly redistributed absorption bands than the neutral precursors, consistent with their markedly dark colors.[12,22] In the ΔAspectra, pronounced negative features appear near 313 nm for **1** and **3**, and near 320 nm for **2**. For the formamidinate complexes **2** and **3**, additional negative features are observed near 390 nm. These negative signals indicate suppression of the original resonant absorptions (i.e., **I** and **II**) upon oxidation, whereas the newly emerged intense bands at displaced wavelengths are consistent with formation of dressed molecule-field states. Within this framework, we interpret the absorption spectra of the cationic complexes in terms of polaritonic transitions arising from selective coupling between the molecular excitations and the intrinsic scattering modes.

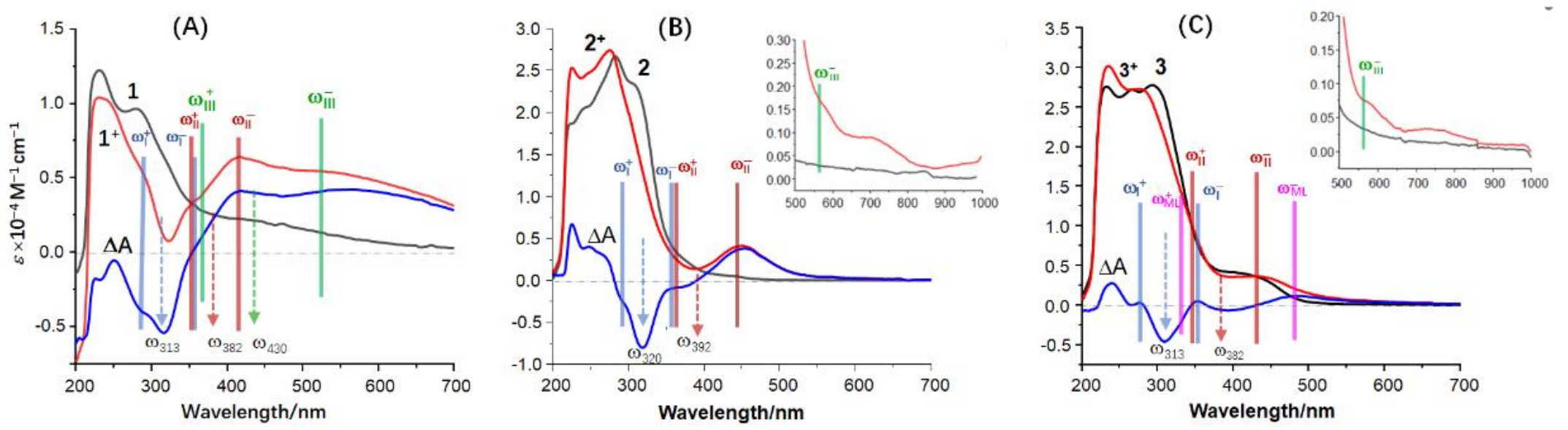

**Figure 3. UV–visible absorption spectra showing mode-selective coupling of molecular excitations to the intrinsic scattering field in the $Mo_2$ complexes.** For each system, the spectra of the neutral complex (black) and the one-electron-oxidized complex (red) are shown together with the corresponding difference spectrum, ΔA (blue). The dashed colored lines indicate the intrinsic photonic modes whose

energies match the relevant molecular transitions, and the solid-colored lines mark the assigned pairs of polaritonic branches. (A) $Mo_2(O_2CCH_3)_4$ (**1**, **1⁺**, and $\Delta$A). (B) $Mo_2(DAniF)_4$ (**2**, **2⁺**, and $\Delta$A). (C) $Mo_2(DAniF)_3(O_2CC_6H_5)$ (**3**, **3⁺**, and $\Delta$A). The paired absorptions are assigned to polaritonic transitions associated with mode-selective coupling of the LMCT excitations **I** and **II**, and of the MLCT excitation $\delta \rightarrow \pi^*$(ph)(pink), as illustrated in Figure 2.

In this $Mo_2$ system, where the emitter and resonator are integrated within a single molecular identity, strong exciton-photon coupling is proposed to be driven by the intrinsic scattering field generated by the dimetal unit.[12,15] Under steady-state spectroscopic conditions, one of the intrinsic scattering modes,[14] $\omega_L$ selectively interacts with a molecular excitation $\omega_0$ when $\omega_L \approx \omega_0$. Within this multiphoton coupling picture, three dressed-state features are expected at $\omega_0$ and $\omega^{\pm} = \omega_0 \pm \Omega$, analogous to a Mollow-type triplet.[15,2627] Thepectrum show the sidebands appearing as the dominant spectral signatures because the fluorescence at resonance is coupled with the excitation.[15,28,29] This behavior differs from vacuum Rabi splitting in a single-photon process, which yields two components $P_{\pm} = \omega_0 \pm \Omega/2$. For complex **1**, the weak band **I** appears as a shoulder near 313 nm in the neutral spectrum (Figure 3A, black), consistent with the deep negative feature in the corresponding $\Delta A$ spectrum. In the spectrum of **1⁺** (Figure 3A, red), two absorption bands are observed at 283 and 350 nm (Table S1), approximately symmetrically distributed about 313 nm. This position corresponds to mode $\omega_{313}$, previously assigned to the blue Mollow sideband ($\omega_0 + N\Omega'$), with $\omega_0 = 26178\ \text{cm}^{-1}$, $\Omega' = 1450\ \text{cm}^{-1}$, and $N = 4$).[14] We therefore assign these two bands to the upper and lower polaritonic branches, $\omega_I^+$ and $\omega_I^-$, respectively. Their separation corresponds to $2\Omega = 6272\ \text{cm}^{-1}$, giving a coupling strength of $\Omega = 3136\ \text{cm}^{-1}$. The normalized coupling ratio, estimated as $\eta = \Omega/(2\omega_0)$, is approximately 0.05, consistent with the strong-coupling regime for this hybrid system. Similar spectral behavior is observed for **2⁺** and **3⁺**. For **2⁺**, the $\Delta$A spectrum shows a deep minimum near 320 nm, flanked by bands at 294 and 352 nm that are assigned to $\omega_I^+$ and $\omega_I^-$, respectively. This wavelength (320 nm) matches the coherent mode $\omega_{320}$, which differs from $\omega_{313}$ by phase shift of $\Omega/2$ ($\Omega = 1380\ \text{cm}^{-1}$) or $\pi$. For **3⁺**, the corresponding funnel-shaped feature remains centered near 313 nm, with polaritonic branches at 278 and 357 nm. Together, these results support the assignment of absorptions for the oxidized complexes to polaritonic transitions between the dressed states arising from mode-selective coupling of transition **I**.

The behavior of transition **II** further supports the mode-selective coupling picture. For the neutral complexes, this excitation near 390 nm is not clearly resolved in the room-temperature absorption spectra because of substantial spectral overlap (Figures 3A–3C, black). For the oxidized species **2**$^+$ and **3**$^+$, however, shallow negative features appear near 392 nm in the $\Delta$A spectra, indicating suppression of the corresponding resonant absorption upon coupling to the intrinsic scattering mode $\omega_{392}$, developed by radiative decay of the lower Rabi-splitting state (P−) for the single molecules.[14] For **2**$^+$, the intense band near 427 nm is assigned to the lower polaritonic branch $\omega_{II}^-$, while the upper branch $\omega_{\mathrm{II}}^+$ is placed at 365 nm, where it overlaps with $\omega_I^-$ (Figure 3B). A similar analysis applies to **3**$^+$, for which the polaritonic branches associated with transition **II** are assigned to absorptions at 345 and 430 nm (Figure 3C). This assignment to the coupled states explains the displacements of the resonance and the substantial gain of the absorptions for the oxidized complex. For complex **1**, transition **II** is expected at higher energy because replacement of N donors by O donors raises the corresponding electronic excitation energy.[24,31] Consistent with this trend, **1**$^+$ does not show a 392 nm valley in the $\Delta$A spectrum; instead, its coupled state results from interaction with a higher-energy intrinsic mode, denoted here as $\omega_{382}$ or $\omega_0$ of the $Mo_2$ resonator.[14] The pronounced absorption at 416 nm is then assigned to the lower polaritonic branch $\omega_{II}^-$, higher in energy than that for **2**$^+$, while the upper branch is expected near 353 nm and overlaps with $\omega_I^-$ (Figure 3A). The derived coupling strength is of the same order of magnitudes as those obtained for **2**$^+$ and **3**$^+$, supporting a common coupling mechanism across this family of oxidized $Mo_2$ complexes.

Complex **3** provides an especially informative case because its broad absorption envelope from approximately 380 to 480 nm contains contributions from transition **II**, the **MLCT** excitation, and transition **III** near 446 nm (Figure 3C, black).[12] Upon one-electron oxidation, the absorption of **3**$^+$ extends further to lower energy, reaching about 500 nm. This additional low-energy intensity is most reasonably attributed to coupling of the MLCT excitation, which introduces an extra two-level manifold not present in complexes **1** and **2**. Within the coupling picture developed here, the band at 483 nm is assigned to the lower polaritonic branch of the MLCT excitation, $\omega_{\mathrm{ML}}^{-}$(Figure 3C, pink), while the corresponding upper branch, $\omega_{\mathrm{ML}}^{+}$, is placed at 330 nm through near-resonant coupling to the $\omega_{392}$scattering mode. This assignment is consistent with prior fluorescence and femtosecond transient measurements that detected the same low-energy feature.[12] The presence of $\omega_{\mathrm{ML}}^{+}$near 330 nm is manifested by the reduced depth and blue-shifted shape of the band-**I** absorption valley in the ΔA spectrum of **3**$^{+}$ relative to that of **2**$^{+}$. From the observed splitting, the coupling strength for the MLCT excitation is estimated to be about 4800 cm$^{-1}$, corresponding to a normalized coupling ratio of $\eta \approx 0.1$, which places this interaction in the ultrastrong-coupling regime. The increased coupling rate for the MLCT excitation in **3**$^{+}$ is obviously due to its large transition dipole moment. The broad negative feature around 392 nm in the ΔAspectrum is therefore assigned to simultaneous bleaching of transition **II** and the MLCT excitation, both of which couple near-resonantly to the same intrinsic scattering mode, $\omega_{392}$.

The $\delta \rightarrow \delta^{*}$transition (**III**) provides a further test of the mode-selective coupling model. For complex **1**, this transition occurs near 435 nm,[21] higher in energy than in the formamidinate analogues **2** and **3** ($\sim 445$ nm)[12,22] with a N-donor coordiantion shell. In the $\Delta A$ spectrum of **1**$^{+}$, a shallow minimum appears near 470 nm, accompanied by a broad absorption extending from about 450 to 700 nm with a maximum near 520 nm (Figure 3A). Similar broad low-energy absroption spectrum was also observed for the singly-oxidized $Ni_2$ complex,[30] which has been attributed to optical transitions of the polaritonic states.[15] For **1**$^{+}$, these features are consistent with coupling of transition **III** to the intrinsic scattering mode $\omega_{430}$.[14] Similarly, a pair of absorptions has been observed near 365 and 530 nm for oxidized dimolybdenum complex $[\mathrm{Mo_2(TiPB)_4}]^{+}$(TiPB = 2,4,6-triisopropylphenyl carboxylate).[31] Rather than assigning these features as ordinary electronic transitions, we interpret them as the upper and lower polaritonic branches, $\omega_{III}^{+}$and $\omega_{III}^{-}$, centered about the $\delta \rightarrow \delta^{*}$ resonance.

Earlier studies[24,32] on **1** and $Mo_2(O_2CCF_3)_4$ showed broad emission around 550 nm with vibrational progression of the mode ($\nu_{\text{Mo-Mo}} \approx 400\ \text{cm}^{-1}$). The hyperfine structure of the low-energy emission retains the Mo–Mo stretching character of $Mo_2$-based electronic ground state in accordance with the Franck-Condon principle, indicating an inherited feature of the dressed states arising from the parent $\delta \rightarrow \delta^*$ excitation. Therefore, the broad absorption at 520 nm for **1**$^+$ is assigned to $\omega_{III}^-$, and the corresponding upper branch is placed near 365 nm (Figure 3A), giving a coupling strength of about 4000 cm$^{-1}$. The oxidized formamidinato complexes **2**$^+$ and **3**$^+$ show very weak absorption near 565 nm (Figures 3B and 3C, insets), with the $\delta \rightarrow \delta^*$transition shiftting to about 445 nm, consistent with previously noted for related systems.[22] Even so, the presence of this low-energy feature, together with the additional weak absorptions at 700–800 nm that are absent in the neutral precursors, suggests that oxidation opens further low-energy polaritonic channels in the coupled $Mo_2$ systems.

**The molecular excitons for the neutral and cationic complex systems.** The observation that the two-level excitations in the cationic $Mo_2$ complexes, rather than in the neutral precursors, exhibit the clearest signatures of coupling to the scattering field can be rationalized in terms of the nature and binding strength of the corresponding molecular excitons. For the neutral complex, the exciton binding energy may be introduced as the energy required to separate the optically generated electron-hole pair into noninteracting charge carriers within the molecular framework.[33] This process may be represented formally as

$$2[Mo_2] + E_b \rightarrow [Mo_2]^+ + [Mo_2]^- \qquad (1)$$

Figure 4A schematically compares the MLCT excitations of the neutral and oxidized species, which generate the corresponding excitons $[Mo_2]^*$ and $[Mo_2]^{+*}$. The binding energy $E_b$ is therefore expressed as the difference between the fundamental gap $E_g$ and the optical gap $E_g^{\text{opt}}$,[34, 35] namely,

$$E_b = E_g - E_g^{\text{opt}} \qquad (2)$$

where $E_g^{\text{opt}}$is given by the optical transition energy, and the energy required to generate the corresponding separated charge carriers is termed as transport energy or fundmental gap $E_g$. In molecular terms, $E_g$ can be written as[36]

$$E_g = IP - EA \qquad (3)$$

where $IP$ and $EA$ are the first ionization potential and electron affinity, respectively. To first approximation, the fundamental gap $E_g$ corresponds to the HOMO-LUMO gap $E_g \approx \Delta E_{H\text{-}L}$, as illustrated in the energy diagram (Figure 4C).[37,38]

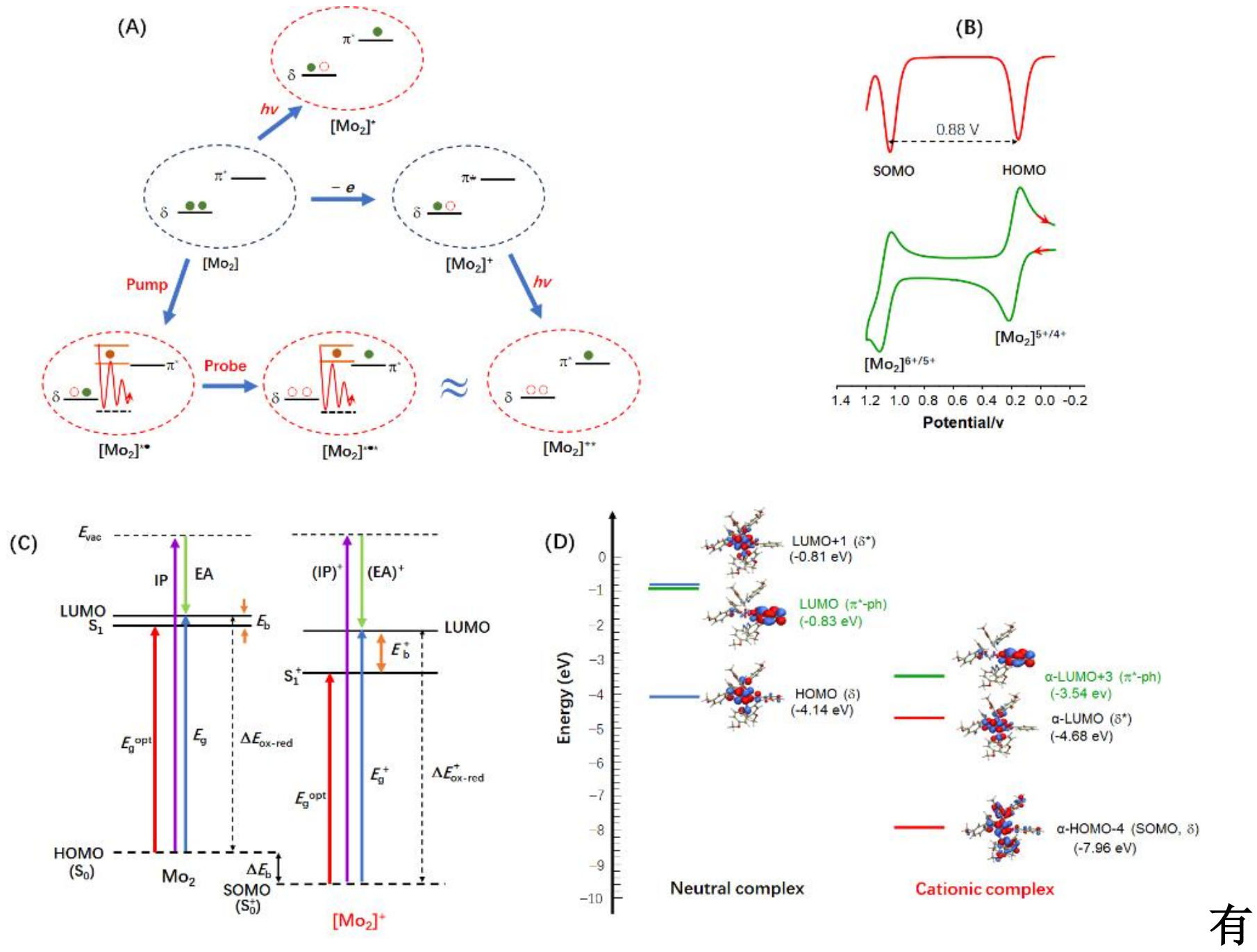

**Figure 4. Schematic and experimental analyses of the excitonic structures of the neutral and cationic $Mo_2$ complex systems.** (A) Illustration of the MLCT-type excitons formed in the neutral complex, the one-electron-oxidized complex, and the corresponding polaritonic excitons under the relevant spectroscopic conditions. (B) Differential-pulse voltammetry (DPV, red) and cyclic voltammetry (CV, green) of $Mo_2(DAniF)_4$ (**2**) showing two successive one-electron oxidation processes associated with the $Mo_2$ center. (C) Energy diagrams illustrating the binding energies of the neutral and cationic $Mo_2$ excitons. (D) Frontier molecular orbitals calculated for **3** and **3⁺** at the time-dependent DFT level, showing the relative changes in the metal-based and ligand-based orbital energies upon one-electron oxidation.

When both redox processes are electrochemically accessible, $E_g$ may also be estimated from the separation between the oxidation and reduction potentials.[34,37] For the two $Mo_2$ systems in different oxidation states, $E_g = E_{ox} - E_{red}$ for the neutral complex and $E_g^+ = E_{ox}^+ - E_{red}^+$ for the cationic complex. Thus, the difference in transport energy between the two systems is therefore

$$\Delta E_g = E_g^+ - E_g = (E^+_{ox} - E^+_{red}) - (E_{ox} - E_{red}) = E^+_{ox} - E_{ox} \quad (5)$$

where the last approximation follows from assuming $E^+_{red} \approx E_{red}$ for reduction processes that place one electron into the ligand-centered $\pi$*(ph) oribital of the neutral and cationic molecules. Here, $E^+_{ox}$ and $E_{ox}$ correspond to the redox potentials for the couples $[Mo_2]^{6+/5+}$ and $[Mo_2]^{5+/4+}$ , respectively. As shown in Figure 4B, the DPV and CV for complex **2** in dichloromethane exhibit two reversible one-electron redox processes associated with the $Mo_2$ center. The first oxidation potential $E_{ox}$ = 0.186 V, corresponding to conversion of $[Mo_2^{4+}]$ to $[Mo_2^{5+}]$, whereas the second oxidation removes the second δ electron with $E^+_{ox}$ = 1.07 V. To first apporoximation, and neglecting solvent-dependent contributions, the potential separation of 0.88 V therefore provides a practical measure of increase in transport gap for the oxidized complex relative to the neutral precursor that is, $\Delta E_g$ according to Eq. 5. If the two systems have similar $E_g^{opt}$, then, Eq. 2 further implies that $\Delta E_b \approx \Delta E_g$. Similar separations of approximately 0.8–1.0 V have been reported for related $Mo_2$ complexes,[39,40] indicating that this energetic effect is relateively insensitive to ligand enviroment.

The electrochemically determined $\Delta E_g$ value may also be viewed as an approximation to the energy difference between the SOMO of the cationic complex and the HOMO of the neutral, $\Delta E_{SOMO-HOMO}$, if both systems are referenced to a comparable ligand acceptor level. Figure 4D shows the frontier molecular orbitals calculated for **3** and **3**$^+$ at the time-dependent DFT level (see Supplementary Materials). For the neutral complex **3**, the HOMO is the $Mo_2$-based $\delta$ orbital and the LUMO is the ligand-centered $\pi^*$(ph)orbital, which lies below the metal-based $\delta^*$ level. Upon one-electron oxidation, the energy levels of the electronic MLCT manifolds are lowered substantially, whereas the $\delta \rightarrow \delta^*$ excitation is affected to a lesser extent. In particular, the SOMO of **3**$^+$ lies markedly below the HOMO of **3**, while the ligand $\pi^*$(ph) level is lowered by a smaller amount. The calculations therefore predict an apprecibale increase of the MLCT transition energy for the cationic complex, in sharp contrast to the experimental observed red shift of MLCT absorption (Figure 3C). This discripancy provides additional support for the conclusion that the optical responses of the oxidized complexes is governed by light-coupling behavior rather than by ordinary electronic structure shifts alone. The larger stabilization of the metal-based orbital relative to the ligand acceptor level gives an estimated $\Delta E_{\mathrm{SOMO-HOMO}}$ of about 1.11 eV for the MLCT systems of **3**/**3**$^{+}$, and a similar value of 1.22 eV was obtained for **2**/**2**$^{+}$ (see Supplementary Materials). Although the calculations likely overestimate the absolute lowering of the metal-based level because the SOMO retains appreciable ligand character (Figure 4D), the computed trend agrees well with the electrochemical estimate of an energy decrease of approximately 0.9 eV, indicating a notably large binding energy for the oxidized species.[33,34,36] Taken together, the electrochemical and computational results support the conclusion that one-electron oxidation substantially increases the binding energy of the MLCT exciton. This more strongly bonded exciton is expected to possess larger oscillator strength and to couple more effectively to the intrinsic scattering field,[41,42,43] thereby accounting for the pronounced Rabi-type splittings observed for the cationic $Mo_2$ complexes at room temperature, whereas the neutral precursors do not show comparably resolved coupling signatures.

**Polariton transitions observed in resonance fluorescence and ultrafast transient spectra.** Complex **3** provides further evidence for the proposed polaritonic assignments through both photoluminescence and ultrafast transient spectroscopy. In the photoluminescence spectra (Figure 5A), excitation at 300–310 nm produces emission peaks at 3655 and 475 nm, whereas excitation at 420-430 nm gives rise to an intense band at 565 nm. The positions of 365 nm and 565 nm are comparable to the absorption wavelengths for **1**$^{+}$ (Figure 3A), which have been assigned to the upper and lower polaritonic branches associated with the $\delta \rightarrow \delta^{*}$ excitation, $\omega_{III}^{+}$ and $\omega_{III}^{-}$, respectively. The 565 nm transition is also observed in absorption spectra for **2**$^{+}$ and **3**$^{+}$ (Figures 3B and 3C, insets). However, these two emission bands are centered about 443 nm, indicating resonant coupling of the $\delta \rightarrow \delta^{*}$ transition to the photonic mode $\omega_{443}$ evolving from the Rabi sideband at ($\omega_0 - N\sqrt{N}\Omega/2$) ($\omega_0 = 26178^{-1}$, $\Omega = 1380$ cm$^{-1}$ and $N = 3$),[14] as a result of resonance III shifting to lower energy for the formamidinate analogues. The intermediate band at 475 nm has been attributed to the lower sideband of the coupled MLCT system, $\omega_{\text{ML}}^{-}$. The same low-energy MLCT-related polaritonic feature emerges in the TA spectra with a delay time of about 0.4 ps (Figure 5B), consistent with our earlier transient study.[12] The observation of the same transitions in independent spectroscopic measurements strengthens the interpretation that the absorption, fluorescence, and transient signals all originate from common molecule-field hybrid states. The correspondence between the steady-state absorption spectrum of the oxidized complex and the emission and TA absorption spectra of the neutral molecule highlights an important mechanistic question: why does the cationic complex show the clearest coupling signatures under steady-state conditions, whereas consistent polaritonic absorptions are observed for the neutral species with laser pumping? This issue can be understood by considering the different excitonic configurations generated in the related external optical environments.

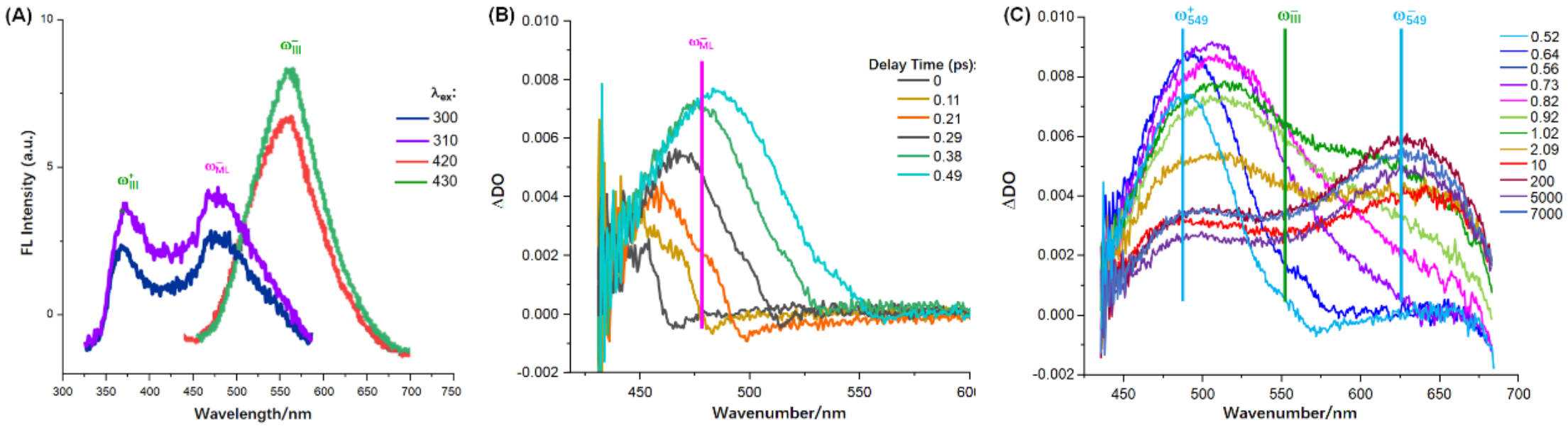

**Figure 5. Resonance fluorescence and ultrafast transient spectra of 3, showing polaritonic transitions associated with the coupled MLCT and $\delta \rightarrow \delta^*$ manifolds.** (A) esonance fluorescence spectra of **3** showing the polaritonic emissions assigned to $\omega_{III}^{+}$ and $\omega_{III}^{-}$ (green), associated with the $\delta \rightarrow \delta^*$ excitation, together with the intermediate band $\omega_{\text{ML}}^{-}$(pink), assigned to the lower sideband of the coupled MLCT system. (B) Transient absorption spectra showing the emergence of the $\omega_{\text{ML}}^{-}$ feature at a delay time of approximately 0.4 ps. (C) Transient spectra showing secondary splitting of the low-energy polaritonic state near 550 nm through coupling to a coherent scattering mode, giving rise to the paired branches $\omega_{550}^{+}$ and $\omega_{550}^{-}$. The

long-lived response of this hybrid state indicates an extended coherent lifetime in the coupled $Mo_2$ system.

Under the TA spectroscopic conditions, laser pumping first excites the neutral molecule to generate an excited (dressed) state with a singly occupied $\delta$ orbital, denoted here as $[Mo_2]^{*\bullet}$. With respect to the MLCT manifold, this intermediate state is electronically analogous to the ground state of the oxidized complex with a singly occupied $\delta$ orbital (Figure 4A). Subsequent interaction with the probe light promotes this species to a higher excitonic state, $[Mo_2]^{*\bullet *}$, in which two holes reside in the $\delta$ manifold and the excited electron occupies the $\pi^*$ (ph) orbital. This transiently generated exciton closely resembles the MLCT exciton of the cationic complex in both electronic configuration and binding character. As a result, the two systems couple to the same intrinsic scattering mode and exhibit closely corresponding polaritonic absorption features under different optical conditions. In this way, optical experiments with laser pumping, including TA absorption and steady-state fluorescence spectroscopies, provide a dynamic route to the same coupled-state manifold that is accessed directly in the oxidized complex under steady-state conditions.

At longer delay times, the transient spectra evolve into two broad absorption bands with maxima at 490 and 624 nm, centered at approximately 549 nm (Figure 5C). This central energy coincides with the $\omega_{III}^{-}$ feature identified at 562 nm in both the absorption spectrum of $\mathbf{3}^+$ (Figure 3C, inset) and the photoluminescence spectrum of **3** (Figure 5A). We therefore assign these two transient bands to a secondary splitting of the low-energy polariton derived from the $\delta \rightarrow \delta^*$ excitation, arising from resonant coupling of the $\omega_{III}^{-}$ state to a coherent scattering mode near 549 nm. This $\omega_{549}$ mode in the scattering field is defined by its displacement from the Mollow sideband at ($\omega_0$ − 5 × $\Omega_1'$) by 725 cm$^{-1}$ or $\Omega_1'/2$, where $\omega_0$ = 26178 cm$^{-1}$, $\Omega_1'$ = 1450 cm$^{-1}$ and $N$ = 5.[14] Within this interpretation, the two transient absorptions correspond to the upper and lower branches of the driven polariton ($\omega_{III}^{-}$),[44] denoted $\omega_{562}^{+}$and $\omega_{562}^{-}$, separated by 4383 cm$^{-1}$. The coupling strength $\Omega_{562}'$ is then determined to be 2191 cm$^{-1}$. Notably, the two branches are asymmetric in intensity, with the lower-energy component weaker than the higher-energy one. Such asymmetry is consistent with a non-resonant multiphoton coupling process with finite detuning $\Delta$, in which absorption and gain contributions to the two branches are unequal.[26,45] For this coupling system, with $\Delta = \omega_L(549\text{ nm}) - \omega_0(562\text{ nm}) >> 0$, the three-photon model predicts that the high-energy polaritonic transition involves photonic emission at $\omega_L + \Omega'_{562}$ , whereas the low-energy polaritonic transition results in an absorption band at $\omega_L - \Omega'_{562}$, as observed in the TA spectra (Figure 5C). It is worthwhile to note that this light-hybrid δ→δ* system exhibits exceptionally long coherent time up to the nanosecond scale (Figure 5C). This long-time coherence for the driven polariton is in remarkable agreement with the long-lived emission (2 μs) at 550 nm for $Mo_2$ complexes,[21] which is assigned to the low-energy polarion branch ($\omega_{III}^-$) in this study. Therefore, observation of this polariton excitation in the TA spectra further supports the spectral assignments to the coupled δ→δ* excitation and its sideband excitation. The transient absorption spectra therefore support the view that the observed sidebands arise from a driven polaritonic process rather than from simple vacuum Rabi splitting in a single-photon regime. More broadly, these results indicate that, in the $Mo_2$ system, not only the primary electronic excitation but also an already formed polaritonic state[44] can undergo further scattering-driven splitting, thereby extending the hierarchy of hybrid light-matter states accessible in the molecular resonator.

**Conclusion**

In summary, this work shows that quadruply bonded $Mo_2$ complexes provide a chemically defined molecular platform in which intrinsic quantized scattering modes can resonantly or near-resonantly couple to selected two-level excitations under ambient conditions. The LMCT, MLCT, and $\delta \rightarrow \delta^*$ manifolds exhibit distinct and mode-selective polaritonic responses, as evidenced by the correlated features observed in the steady-state absorption, resonance fluorescence, and ultrafast transient spectra. The spectral valleys at the original resonances, the emergence of split sidebands, and the observation of low-energy long-lived hybrid states together support the formation of dressed molecule-field states in these $Mo_2$ systems.

A central result of this study is that the clearest coupling signatures are observed for the singly oxidized complexes. Electrochemical and orbital analyses indicate that one-electron oxidation substantially increases the binding strength of the relevant excitons by stabilizing the metal-based frontier orbital relative to the ligand acceptor level. This energetic change provides a physical basis for the stronger oscillator strength and the more clearly resolved scattering-driven exciton-photon coupling observed in the cationic species. In this way, the distinct spectral behavior of the neutral and oxidized complexes is linked to a fundamental change in exciton character, rather than to a simple redox-induced shift of electronic transitions.

Taken together, these results strengthen the emerging picture of bonded dimetal units as integrated molecular emitter-resonator quantum systems. In the $Mo_2$ complexes, the intrinsic scattering field is not a passive radiative by-product, but an active quantized field that selectively reorganizes molecular excitations into polaritonic states. The chemically tunable manifold of $Mo_2$ excitations, together with the sensitivity of the coupling behavior to oxidation state and ligand environment, makes this system a useful platform for exploring molecular quantum optics and, more broadly, for investigating how polaritonic interactions may influence excitation dynamics, energy flow, and charge-transfer processes in complex molecular systems.

**Data availability**

All data are available in the manuscript or the supplementary materials. The spectra raw data have been deposited at Mendeley (Mendeley Data: doi: 10.17632/mxbg5b62hp.3 and are publicly available as of the date of publication. Additional information about the data will be made available from the corresponding author upon reasonable request.

**Acknowledgments**

We acknowledge the primary financial support from the National Natural Science Foundation of China (22171107, 21971088, 21371074), Natural Science Foundation of Guangdong Province (2018A030313894), Jinan University, and the Fundamental Research Funds for the Central Universities.

## Author Contribution

C.Y.L. conceived this project and designed the experiments and worked on the manuscript. M.M. carried out the major experimental work and theoretical simulations, and prepared the Supplementary Information. Y.N.T., Z.C.H. and Y.L.Z. involved in the spectroscopic data analysis and assisted in manuscript preparation.

Competing interests: The authors declare no conflict of interest.

---